\documentstyle[epsfig]{mn}

\newif\ifAMStwofonts

\newcommand{\half}{{\textstyle \frac{1}{2}}}
\newcommand{\rhodot}{\dot{\rho}}
\newcommand{\rdot}{\dot{r}}
\newcommand{\deriv}[2]{\frac{\partial #1}{\partial #2}}
\newcommand{\et}[1]{e^{\mbox{\footnotesize $#1$}}}
\newcommand{\phidot}{\dot{\phi}}
\newcommand{\nn}{\nonumber}
\renewcommand{\r}[1]{(\ref{#1})}

\ifoldfss
  \ifCUPmtlplainloaded \else
    \NewTextAlphabet{textbfit} {cmbxti10} {}
    \NewTextAlphabet{textbfss} {cmssbx10} {}
    \NewMathAlphabet{mathbfit} {cmbxti10} {} 
    \NewMathAlphabet{mathbfss} {cmssbx10} {} 
  \fi
  \ifAMStwofonts
    \ifCUPmtlplainloaded \else
      \NewSymbolFont{upmath} {eurm10}
      \NewSymbolFont{AMSa} {msam10}
      \NewMathSymbol{\upi}     {0}{upmath}{19}
      \NewMathSymbol{\umu}     {0}{upmath}{16}
      \NewMathSymbol{\upartial}{0}{upmath}{40}
      \NewMathSymbol{\leqslant}{3}{AMSa}{36}
      \NewMathSymbol{\geqslant}{3}{AMSa}{3E}

      \let\leq=\leqslant 
       
    \fi
  \fi
\fi 

\ifnfssone
  \newmathalphabet{\mathit}
  \addtoversion{normal}{\mathit}{cmr}{m}{it}
  \addtoversion{bold}{\mathit}{cmr}{bx}{it}
  \newmathalphabet{\mathbfit} 
  \addtoversion{normal}{\mathbfit}{cmr}{bx}{it}
  \addtoversion{bold}{\mathbfit}{cmr}{bx}{it}
  \newmathalphabet{\mathbfss} 
  \addtoversion{normal}{\mathbfss}{cmss}{bx}{n}
  \addtoversion{bold}{\mathbfss}{cmss}{bx}{n}
  \ifAMStwofonts
    \ifCUPmtlplainloaded \else
      \UseAMStwoboldmath
      \makeatletter
      \new@mathgroup\upmath@group
      \define@mathgroup\mv@normal\upmath@group{eur}{m}{n}
      \define@mathgroup\mv@bold\upmath@group{eur}{b}{n}
      \edef\UPM{\hexnumber\upmath@group}
      \new@mathgroup\amsa@group
      \define@mathgroup\mv@normal\amsa@group{msa}{m}{n}
      \define@mathgroup\mv@bold\amsa@group{msa}{m}{n}
      \edef\AMSa{\hexnumber\amsa@group}
      \makeatother
      \mathchardef\upi="0\UPM19
      \mathchardef\umu="0\UPM16
      \mathchardef\upartial="0\UPM40
      \mathchardef\leqslant="3\AMSa36
      \mathchardef\geqslant="3\AMSa3E

      \let\leq=\leqslant 

    \fi
  \fi
\fi 

\ifnfsstwo
  \DeclareMathAlphabet{\mathbfit}{OT1}{cmr}{bx}{it}
  \SetMathAlphabet\mathbfit{bold}{OT1}{cmr}{bx}{it}
  \DeclareMathAlphabet{\mathbfss}{OT1}{cmss}{bx}{n}
  \SetMathAlphabet\mathbfss{bold}{OT1}{cmss}{bx}{n}
  \ifAMStwofonts
    \ifCUPmtlplainloaded \else
      \DeclareSymbolFont{UPM}{U}{eur}{m}{n}
      \SetSymbolFont{UPM}{bold}{U}{eur}{b}{n}
      \DeclareSymbolFont{AMSa}{U}{msa}{m}{n}
      \DeclareMathSymbol{\upi}{0}{UPM}{"19}
      \DeclareMathSymbol{\umu}{0}{UPM}{"16}
      \DeclareMathSymbol{\upartial}{0}{UPM}{"40}
      \DeclareMathSymbol{\leqslant}{3}{AMSa}{"36}
      \DeclareMathSymbol{\geqslant}{3}{AMSa}{"3E}

      \let\leq=\leqslant 

    \fi
  \fi
\fi 

\ifCUPmtlplainloaded \else
  \ifAMStwofonts \else 
    \def\upi{\pi}
    \def\umu{\mu}
    \def\upartial{\partial}
  \fi
\fi

\title[Microwave background anisotropies and nonlinear structures I ]
  {Microwave background anisotropies and nonlinear structures~~I.
  Improved theoretical models}   
\author[A.N.~Lasenby et al.]
  {A.N.~Lasenby\thanks{Email: anthony@mrao.cam.ac.uk}, C.J.L.~Doran,
  M.P.~Hobson, Y.~Dabrowski and A.D. Challinor  \\ 
  Mullard Radio Astronomy Observatory, Cavendish Laboratory,
Madingley Road, Cambridge CB3 0HE, UK
}

\date{Accepted ???. Received ???; in original form \today} 
\pagerange{\pageref{firstpage}--\pageref{lastpage}}
\pubyear{1997}

\begin{document}

\label{firstpage}

\maketitle

\begin{abstract}
A new method is proposed for modelling spherically symmetric
inhomogeneities in the Universe.  The inhomogeneities have finite size
and are compensated, so they do not exert any measurable gravitational
force beyond their boundary.  The region exterior to the perturbation
is represented by a Friedmann-Robertson-Walker (FRW) Universe, which
we use to study the anisotropy in the cosmic microwave background
(CMB) induced by the cluster.  All calculations are performed in a
single, global coordinate system, with nonlinear gravitational effects
fully incorporated.  An advantage of the gauge choices employed here
is that the resultant equations are essentially Newtonian in form.
Examination of the problem of specifying initial data shows that the
new model presented here has many advantages over `Swiss cheese' and
other models.  Numerical implementation of the equations derived here
is described in a subsequent paper.
\end{abstract}


\section{Introduction}

Many attempts have been made to model secondary anisotropies in the
cosmic microwave background (CMB) due to galaxy clusters.  The most
widely employed approach is to use a linearised set of equations to
incorporate gravitational effects 
\cite{land-fields,mss90,ms90,chod91,pyn96}.
This has the advantage that one can deal with quite general matter
perturbations, avoiding the restrictions of spherical symmetry imposed
by most other methods.  Doubts always remain, however, over the
accuracy of calculations performed in the linearised theory, and it is
only through comparison with models incorporating full, nonlinear
effects that any errors can be reliably computed.

The earliest attempts to calculate anisotropies in the full, nonlinear
theory employed `Swiss cheese' models 
\cite{rees68,dyer76,Kais82,nott82,nott82II,nott83,nott84}.
In these models a collapsing FRW Universe is surrounded by a
compensating vacuum region, which then matches onto an exterior
expanding FRW Universe.  The compensating region ensures that the
perturbation has no net gravitational effect on the exterior Universe.
These models have two key advantages: analytic calculations can be
performed in the fully nonlinear regime; and all observations can be
modelled in the exterior region, providing clear, unambiguous
predictions of the effect of the cluster.  The main disadvantage of
such models is that the matter distribution is unrealistic, and
appears to overestimate the effect of the cluster~\cite{quil95}.

More realistic models can be constructed by working with an arbitrary
density profile while restricting to spherical symmetry and ignoring
pressure~\cite{pan92,arn93,saez93,arn94,ful94,quil97}.
Spherically symmetric models with
vanishing pressure have implicit analytic solutions in General
Relativity, one simple form of which is provided by the Tolman-Bondi
solution~\cite{tol34,bon47}. Models based on this solution provide a
more reliable method for computing anisotropies due to clusters or
voids, and it is a hybrid of this scheme that we discuss here.
Previous work with Tolman-Bondi models has considered density and
velocity perturbations, but compensating effects are not usually
included.  (One exception to this is the `type I' class of models
considered by Panek~\shortcite{pan92}, but these are only compensated at
infinity.)  The fact that observers comoving with the fluid are not in
a region modelled by a homogeneous FRW cosmology makes it harder to
discuss perturbations, since these observers see a dipole anisotropy
in the CMB, which is essentially an artefact of the model.  

One difficulty in constructing compensated models is that the initial
density and velocity profiles must be chosen in such a way that
streamline crossing is avoided.  Without this, shock fronts form and
one would have to include pressure to produce a realistic model.  For
example, if one just perturbs the density profile without perturbing
the velocity field, then streamline crossing is inevitable in any
compensated model other than those of `Swiss cheese' type (where the
problem would only occur in the vacuum region).

Here we discuss a family of models which avoid the problem of
streamline crossing in a very simple manner, while keeping the density
profile compensated and realistic.  The initial perturbation is of
finite extent and is controlled by two physical parameters, one
describing the magnitude of the perturbation and the other its linear
extent.  The perturbation drops to zero at a finite radius, beyond
which the Universe is described by an FRW model.  A third parameter,
$m$, controls the degree of the polynomial describing the perturbation
and can be adjusted to produce the desired structure in the nonlinear
regime.  The two parameters controlling the perturbation and the
further two parameters describing the external universe uniquely
determine the form of the velocity perturbation for each polynomial
degree.  This is achieved by matching the first $m$ derivatives at the
centre and edge of the perturbation, leaving a polynomial of degree
$2m+1$.  From the velocity profile the density perturbation is
uniquely fixed by the constraints that there are no decaying modes
present and that the density distribution is compensated.  In these
models density compensation holds at all later times and the velocity
field evolves in a way that avoids streamline crossing.  The fact that
the initial density and velocity profiles are described by simple
polynomials also simplifies the task of carrying out accurate
numerical work.

In this paper we describe the theoretical aspects of our model and
provide a detailed comparison with other work.  In Section~\ref{TEQ}
we introduce a novel way of treating the field equations for
spherically symmetric dust, which arose from a new, gauge-theoretic
treatment of gravity presented elsewhere (Lasenby, Doran \& Gull
1998).  This method works with a single global time coordinate which
reduces the equations to an essentially Newtonian form.  The insights
provided by this gauge lead naturally to the idea of specifying
initial conditions on a surface of constant `Newtonian' time, and we
show how to parameterise the matter streamlines to take advantage of
this possibility.

In Section~\ref{S-PH} the equations for photon propagation are
presented, and we derive a simple formula for the temperature
decrement induced in the CMB by the cluster.  We also derive a new
formula for density evolution along a streamline.  This considerably
simplifies the task of computing the density profile at later times.
The advantages of our gauge choices are further illustrated by
linearising the field equations to produce a relation between the
velocity and density profiles for realistic initial conditions.  In
Section~\ref{S-SCM} we discuss Swiss Cheese models and give a new
formula for the CMB temperature decrement in such models.  We also
consider the strengths and limitations of these models as a prelude to
presenting an improved scheme.

Our improved model is presented in Section~\ref{S-MOD}.  Despite
the simplicity of the model, numerical simulations reveal that the
initial perturbation evolves to produce a density profile which
accurately matches the King profile observed in many galaxy
clusters.  The main limitations of our model concern the
imposition of spherical symmetry and the lack of pressure support
in the evolving cluster. Both assumptions lead to large cluster
infall velocities relative to the CMB, and these are certainly
much larger than the velocities observed in nearby clusters.
However, recent numerical work~\cite{nav96} has suggested that
large infall velocities might occur in evolving clusters at higher
redshifts, for which our model would then be well suited. For
example, the microwave decrements associated with quasar pairs
reported by Jones et al.~\shortcite{jones97} and Richards et
al.~\shortcite{rich97} may be due to combined Sunyaev-Zel'dovich
and gravitational effects from high redshift clusters. This is
discussed further in an accompanying paper~\cite{DDGHL-II}, where
the theoretical model for the gravitational effect described here
is implemented numerically. Some results of this analysis are
briefly discussed in the concluding section.  Natural units
$G=c=\hbar=1$ are employed throughout.

\section{The Field Equations and the Newtonian Gauge}
\label{TEQ}

The equations used here were first derived by Lasenby, Doran \&
Gull~\shortcite{DGL-erice} using a new, gauge-theoretic approach to
gravity.  This approach employed `geometric algebra', the mathematical
language which seems to capture best the nature of the gauge fields
and the structure of the equations.  Despite the different conceptual
foundations of the theory developed by Lasenby et
al.~\shortcite{DGL-grav}, the predictions of the theory agree with
general relativity for a wide range of phenomena, and for the systems
studied here the two theories agree exactly.  So as not to alienate
readers unfamiliar with geometric algebra, we have not employed it in
this paper, and have instead adopted an approach closer to that of
standard general relativity.  The price of this approach is that some
of the results employed below are quoted without proof from Lasenby et
al.~\shortcite{DGL-grav}.

The system we study in this paper consists of a spherically symmetric
distribution of dust.  Modifications to include either pressure or a
cosmological constant are not considered here.  Lasenby et
al.~\shortcite{DGL-grav} showed that a natural gauge emerges for
the analysis of such systems.  This gauge is global, employing a
single time coordinate $t$ and a radial coordinate $r$.  The time
coordinate $t$ measures the free-fall time for observers comoving with
the fluid.  In inhomogeneous regions the radial coordinate $r$ is
related to the strength of the tidal force defined by the Riemann
tensor.  The key dynamical variables are the density $\rho(t,r)$, a
velocity field $u(t,r)$ and a generalised `boost' factor denoted by
$\Gamma(t,r)$.  (In Lasenby et al.~\shortcite{DGL-grav}
and~\shortcite{DGL-erice}, $\Gamma$ was denoted by $g_1$ and $u$ was
denoted by $g_2$.)  The fields $\Gamma$ and $u$ define the line
element by
\begin{eqnarray}
ds^2 &=& 
\bigl(1-\frac{u^2}{{\Gamma}^2} \bigr) dt^2 + 2 \frac{u}{{\Gamma}^2} dt \,
dr -\frac{1}{{\Gamma}^2} dr^2  \nn \\
& & -r^2 (d\theta^2 + \sin^2\!\theta \, d\phi^2),
\end{eqnarray}
where $\theta$ and $\phi$ are standard polar coordinates,
$0\leq\theta\leq\pi$, $0\leq\theta<2\pi$.  The density
$\rho(t,r)$ defines a total gravitational mass $M(t,r)$ by
\begin{equation}
M(t,r) \equiv \int_0^r 4 \pi s^2 \rho(t,s)\, ds,
\end{equation}
and the variables $\Gamma$, $u$ and $M$ are related by
\begin{equation}
{\Gamma}^2 = 1 - 2M/r + u^2.
\label{defg1}
\end{equation}
If we rewrite this equation in the form
\begin{equation}
\half u^2 - M/r = \half({\Gamma}^2 -1)
\end{equation}
we see that it has a simple, Newtonian interpretation: it is a
Bernoulli equation for zero pressure and non-relativistic energy
$({\Gamma}^2-1)/2$.

The variable $u$ defines the integral curves of the conserved fluid
current by
\begin{equation}
\frac{dr}{dt} = u.
\end{equation}
These integral curves are also matter geodesics, since there is no
pressure present.  The key dynamical equations are update equations
along these fluid streamlines, for which we require the comoving
derivative
\begin{equation}
\frac{D}{Dt} \equiv \frac{\partial}{\partial t} + u
\frac{\partial}{\partial r}.
\end{equation}
The total gravitational mass enclosed, $M$, is conserved along the
fluid streamlines,
\begin{equation}
\frac{DM}{Dt} = 0.
\label{Mcons}
\end{equation}
Implicit here is the assumption that the initial conditions are chosen
to avoid the possibility of streamline crossing.  The Euler equation
in this setup is
\begin{equation} 
\frac{Du}{Dt} = -\frac{M}{r^2}.
\label{Du}
\end{equation}
It follows that
\begin{equation}
\frac{D\Gamma}{Dt} = 0.
\label{Dg1}
\end{equation}

The above set of equations look Newtonian, yet they are fully
consistent with general relativity.  This is the advantage of the
gauge choices made to manipulate the equations into the above form.
The form of the equations led us to name this gauge the `Newtonian'
gauge.  Initial data for the above set of equations can be given in a
variety of forms, but the most natural is to specify the density
$\rho(t_i,r)$ and the velocity $u(t_i,r)$ at some initial time $t_i$.
From these one can calculate $M(t_i,r)$ and $\Gamma(t_i,r)$, which are
then conserved along the streamlines.

To complete the set of basic equations it is useful to introduce the
velocity gradient $H$:
\begin{equation}
H(t,r) \equiv \frac{\partial}{\partial r}  u(t,r).
\end{equation}
From the above equations it is straightforward to show that
\begin{equation}
\frac{D\rho}{Dt} = -(2u/r +H) \rho.
\label{Drho}
\end{equation}
The function $H$ is shown below to play a fundamental role in
determining the temperature fluctuations caused by an infalling
cluster.

The above set of equations have analytic solutions which were first
found by Tolman~\shortcite{tol34} and Bondi~\shortcite{bon47}.  The
difference between the above approach and the Tolman solution is that
Tolman used $M$ as the radial coordinate, instead of $r$.  This has
the disadvantage of hiding the Newtonian nature of the equations,
making it hard to visualise the physics.  The form of the analytic
solution to the above equations depends on the sign of ${\Gamma}^2-1$
on the initial timeslice.  There are three cases to consider:

\subsubsection*{1. ${\Gamma}^2 < 1$}

For this case the matter streamlines (geodesics) are defined by
\begin{eqnarray}
r &=& \frac{M}{1-{\Gamma}^2} (1- \cos\!\eta) \\
t-t_i &=& \frac{M}{(1-{\Gamma}^2)^{3/2}} ( \eta - \sin\!\eta - \eta_i +
\sin\!\eta_i) 
\end{eqnarray}
where $\eta$ parameterises the curve and $\eta_i$ is determined from
the initial value of $r$ at time $t_i$.  The quantities $M$ and $\Gamma$
are constant along the curve.  In order to choose the initial value
of $\eta$ correctly, one needs the further result that
\begin{equation}
u = \frac{M}{r(1-{\Gamma}^2)^{1/2}} \sin\!\eta
\end{equation}
so that $\eta_i$ is uniquely fixed between $0<\eta_i<2\pi$ from a
knowledge of whether the initial velocity is inwards or outwards.  The
case $\eta_i=\pi$ corresponds to starting from rest, as treated
by Lasenby et al.~\shortcite{DGL-grav}.

\subsubsection*{2. ${\Gamma}^2=1$}

This case includes a flat FRW cosmology, and the equations can be
integrated directly to give
\begin{equation}
t - t_i = \frac{2}{3} \frac{(r^{3/2} - {r_i}^{3/2})}{(2M)^{1/2}}.
\end{equation}
The velocity is chosen to be outwards to avoid a singularity forming
instantaneously.

\subsubsection*{3. ${\Gamma}^2 > 1$}

This case includes open cosmologies.  The matter curves are
parameterised as follows:
\begin{eqnarray}
r &=& \frac{M}{{\Gamma}^2-1} (\cosh\!\eta -1) \\
t-t_i &=& \frac{M}{({\Gamma}^2-1)^{3/2}} (\sinh\!\eta-\eta-\sinh\!\eta_i +
\eta_i) 
\end{eqnarray}
and the velocity is given by
\begin{equation}
u = \frac{M}{r({\Gamma}^2-1)^{1/2}} \sinh \eta.
\end{equation}
For this case it is also necessary to start with an initial outward
velocity to avoid streamline crossing.

The above parameterised expressions for the matter streamlines are
more general than those usually presented~\cite{pan92}.  This is
necessary so that the initial data can be specified on a constant $t$
timeslice, which provides better control over the form of the initial
perturbations. The initial time $t_i$ is arbitrary and is usually
either set to zero or chosen so that the present epoch corresponds to
$t=0$.  Initial data is defined by $\rho(t_i,r)$ and $u(t_i,r)$ and
from these the initial values of $M$ and $\Gamma$ are found.  At the
initial point $r(t_i)$ the value of $\eta_i$ is found, and the
resulting curves are plotted parametrically.  Examples of these curves
for collapsing dust are given by Lasenby et al.~\shortcite{DGL-grav}.
The lack of pressure support means that infalling matter collapses to
a singularity in a finite time.  This does not pose any problems,
provided photons passing through the cluster do not encounter a
singularity or horizon.

\section{Physics in the Newtonian Gauge}
\label{S-PH}

There seems little doubt that the Newtonian gauge provides the
cleanest description of gravitational physics for spherically
symmetric systems~\cite{DGL-grav,DGL96-erice} and it is surprising
that it is not more widely used.  Here we review here some key
physical results from the Newtonian gauge perspective.  These results
are employed in later sections to construct numerical models.  A
further illustration of the use of the Newtonian gauge which is
slightly outside the main theme of this paper is contained in
Appendix~\ref{appA}.

\subsection{Cosmology}

While the Newtonian gauge arises naturally in the study of spherically
symmetric systems, it is not immediately apparent that it is valuable
in the study of homogeneous cosmological models.  Since the models we
discuss below contain a homogeneous region outside the perturbed
region, it is useful to see how standard cosmological notions fit into
this scheme (see also Gautreau~\shortcite{gaut84} and Ellis \&
Rothman~\shortcite{ellis93} for similar, more detailed discussions).

In a homogeneous model the density $\rho$ is a function of time $t$
only, so we have
\begin{equation}
M(r,t) = \frac{4}{3} \pi r^3 \rho(t).
\end{equation}
Equations~\r{Mcons} and~\r{Drho} now yield $H=u/r$ and $\rhodot
=-3H\rho$, where the overdot denotes differentiation with respect to
time.  It follows that $H$ is a function of time only for homogeneous
models.

The Euler equation~\r{Du} yields the second cosmological equation,
\begin{equation}
\dot{H} + H^2 = - \frac{4\pi}{3} \rho.
\end{equation}
It is clear now that we can identify $H$ as the Hubble function and
that the Newtonian time parameter agrees with cosmic time in
homogeneous regions of the model.

To complete the set of cosmological equations we note from
equation~\r{defg1} that ${\Gamma}^2$ must be of the form
\begin{equation}
{\Gamma}^2 = 1 + r^2 \chi(t).
\end{equation}
Conservation of $\Gamma$ down the streamlines (equation~\r{Dg1}) shows
that 
\begin{equation}
\dot{\chi} = -2 H(t) \chi,
\end{equation}
hence we have
\begin{equation}
{\Gamma}^2 = 1 - k r^2 \exp\bigl\{ -2 \int^t H(t')\, dt' \bigr\}.
\end{equation}
The Friedmann equations in their standard form (with zero pressure and
cosmological constant) are recovered by introducing the scale factor
$S(t)$ via the definition
\begin{equation}
H(t) = \frac{\dot{S}(t)}{S(t)},
\end{equation}
so that
\begin{equation}
{\Gamma}^2 = 1 - kr^2/S^2,
\end{equation}
with $k=0, \pm 1$ determining the type of cosmology in the usual
manner.

In the Newtonian gauge version of cosmology, particles in the Hubble
flow are pictured as moving radially outwards at a velocity $\rdot =
H(t) r$.  The expansion centre is not a physical feature, since the
model is homogeneous and all gauge-invariant (measurable) quantities
are functions of time only.  One can often take advantage of this fact
by considering observers located at the origin.  This provides a novel
way of visualising some of the less intuitive aspects of FRW
cosmologies~\cite{DGL-grav}.  Along the matter streamlines $r(t)/S(t)$
is constant.  The standard `co-moving' gauge for cosmology is obtained by
introducing a new radial coordinate $R=r/S$, so that the matter
geodesics are lines of constant $R$.

\subsection{Photon Paths and Redshifts}

In studying photon trajectories it is sufficient to just consider
motion in the $\theta=\pi/2$ plane.  We can parameterise the photon
paths in terms of the Newtonian time parameter $t$ (cosmic time for
homogeneous models), so that the path is defined by $r(t)$ and
$\phi(t)$.  The condition that the trajectory is null means that we
can write
\begin{eqnarray}
\frac{dr}{dt} &=& \Gamma \cos\!\chi + u \\
\frac{d\phi}{dt} &=& \frac{\sin\!\chi}{r}
\end{eqnarray}
where $\chi$ is the angle between the photon trajectory and the
cluster centre, as measured by observers comoving with the fluid.  The
geodesic equation is equivalent to the following first-order equation
for $\chi$:
\begin{equation}
\frac{d\chi}{dt} = \sin\!\chi \bigl( -\Gamma/r + (H-u/r) \cos\!\chi
\bigr). 
\end{equation}
The above set of three first-order equations is simpler to implement
numerically than the second-order equations obtained from the direct
approach.  The equations have the further advantage of dealing
directly with the measurable quantities $t$ and $\chi$.  It is typical
of the gauge theory approach that one obtains first-order equations in
the gauge invariant observables such as these.  Initial data for the
trajectory consists of position data $r_i$ and $\phi_i$ and a
direction $\chi_i$.  For most calculations, however, the data is given
in the form of the observer's position and an angle on the sky $\chi$,
and the equations are then run backwards in time to take the photon
back through the cluster.  This is the simplest way to perform lensing
simulations.

The remaining content of the geodesic equation concerns the frequency
$\omega$ in the rest frame of the fluid.  This satisfies the
equation~\cite{DGL-grav} 
\begin{equation}
\frac{d\omega}{dt} = - \omega \bigl(H \cos^2\!\chi + \frac{u}{r}
\sin^2\!\chi \bigr).
\label{redshift}
\end{equation}
It is not hard to show that the above equations imply that the angular
momentum $L=\omega r^2\phidot=-r\omega \sin\!\chi$ is conserved.

For the case of a homogeneous cosmology, the redshift
equation~\r{redshift} reduces to
\begin{equation}
\frac{d\omega}{dt} = - \omega H,
\end{equation}
from which we recover the standard result that $\omega S$ is constant for
a photon travelling in a homogeneous, dust-filled Universe.  This
result shows how the standard predictions of cosmology are recovered
in this unfamiliar gauge.

To calculate the effect of a cluster on the CMB, we write
\begin{equation}
u(t,r)= r H_e(t) + \Delta(t,r),
\end{equation}
where $H_e(t)$ is the Hubble function in the exterior Universe at the
same time $t$, and $\Delta$ is the difference between the equivalent
velocity of the unperturbed Universe and the cluster velocity.  For a
photon passing through the cluster we find that
\begin{equation}
\frac{d}{dt} \ln(\omega S) = -\frac{\partial \Delta}{\partial r} \cos^2\!
\chi - \frac{\Delta}{r} \sin^2\!\chi.
\end{equation}
We therefore define the function
\begin{equation}
\epsilon = \int_{t_1}^{t_2} \! dt \, \bigl( \frac{\partial \Delta}{\partial
r} \cos^2\! \chi + \frac{\Delta}{r} \sin^2\!\chi \bigr),
\end{equation}
where the integral is evaluated along the photon path between the time
the photon enters the cluster ($t_1$) and the time it leaves ($t_2$).
The function $\epsilon$ is small, since the contribution to the integral
from near the cluster centre tends to cancel the contributions from
further out.  The main effect producing a non-zero $\epsilon$ is
essentially the evolution of $\Delta$ with time.

If the photon enters the cluster from the CMB with frequency $\omega_1$
and reemerges with frequency $\omega_2'$ we have
\begin{equation}
\omega_2' = \omega_1 \frac{S(t_1)}{S(t_2)}  \et{-\epsilon}.
\end{equation}
An equivalent unperturbed photon will have frequency
\begin{equation}
\omega_2 = \omega_1 \frac{S(t_1)}{S(t_2)},
\end{equation}
so the physically measurable temperature decrement is
\begin{equation} 
\frac{\Delta T}{T} = \frac{\omega_2' - \omega_2}{\omega_2} = \et{-\epsilon} - 1
\approx -\epsilon.
\end{equation}
This simple result enables easy computation of the effect of the
cluster for various angles on the sky.

\subsection{Evolution of the Physical Variables}

The two key physical variables required at later times are the density
distribution, which should evolve to produce a realistic distribution
for a cluster, and the velocity gradient, which controls the
propagation of photons through the cluster.  One way to calculate
these at later times is to differentiate numerically the mass $M(t,r)$
and velocity, which are easily found on any given streamline.  A
better approach, however, is to construct explicit analytical
formulae for $\rho$ and $H$ on a given streamline, which we now
discuss.

We start with the density profile and employ the relation
\begin{equation}
\rho(t,r) = \left(\deriv{r_i}{r}\right)_{\!t} \frac{{r_i}^2}{r^2}
\rho(r_i),
\end{equation}
which gives $\rho(t,r)$ as a function of the initial density profile
$\rho(r_i)=\rho(t_i,r_i)$ and the derivative of the streamline
starting position $r_i(t,r)$ with respect to $r$.  To calculate this
latter term we use the reciprocity relation
\begin{equation}
\left( \deriv{r_i}{r} \right)_{\!t} 
\left( \deriv{r}{t} \right)_{\!r_i}  
\left( \deriv{t}{r_i} \right)_{\!r}  = -1
\end{equation}
to deduce that
\begin{equation}
\left( \deriv{r_i}{r} \right)_{\!t}^{-1} = - u\left( \deriv{t}{r_i}
\right)_{\!r}.
\end{equation}
The final term can be calculated directly from the parameterised form
of the streamlines.  It is convenient here to introduce the function
\begin{equation}
f(r_i) \equiv \Gamma(t_i,r_i)^2-1.
\end{equation}
With this we find that, for $f\neq 0$,
\begin{eqnarray}
\left( \deriv{r_i}{r} \right)_{\!t}^{-1} 
&=& - \frac{1}{2f(r_i)M(r_i)} \Bigl[ M(r_i) \frac{df(r_i)}{dr_i}
(2r-3ut) \nn \\
& &
- 8 \pi {r_i}^2 \rho(r_i) f(r_i)(r-ut) \Bigr] \nn \\
& & + \frac{uM(r_i)}{f(r_i)u(r_i)} \frac{d}{dr_i}
\left(\frac{r_if(r_i)}{M(r_i)}\right),
\label{denevol}
\end{eqnarray}
where $M(r_i)=M(t_i,r_i)$ is the initial mass function.  With this
result one need only compute $df/dr_i$ from the initial data to find
the value of $\rho$ on a given streamline.  The case of $f=0$ must be
treated separately and yields simply
\begin{equation}
\left( \deriv{r_i}{r} \right)_{\!t}^{-1} = \left( \frac{r_i}{r}
\right)^{1/2} + \frac{ut}{2M(r_i)} 4 \pi {r_i}^2 \rho(r_i).
\end{equation}

The same technique is used to find the velocity gradient on a streamline at
later times.  On differentiating equation~\r{defg1} we find that
\begin{equation}
\deriv{u}{r} = -\frac{M}{r^2 u} + \left( \deriv{r_i}{r} \right)_{\!t}
\left( \frac{1}{r} \frac{dM(r_i)}{dr_i} + \frac{1}{2}
\frac{df(r_i)}{dr_i} \right),
\end{equation}
which again gives $H$ on a streamline in terms of the streamline label
$r_i$ and the initial data.  These formulae are very useful for
carrying out accurate numerical simulations.

\subsection{The Linearised Equations}

We now derive a physical constraint on the initial conditions for our
model by considering evolution in the linear regime.  Linearising the
field equations around a homogeneous background cosmology is a
straightforward exercise in the Newtonian gauge.  We start by
introducing the variables
\begin{eqnarray}
\delta u &\equiv& u - r \bar{H}(t) \\
\delta \rho &\equiv& \rho - \bar{\rho}(t) \\
\delta M &\equiv&  \int_0^r 4 \pi s^2 \, \delta\rho(t,s)\, ds,
\end{eqnarray}
where the barred quantities denote the background homogeneous values.
The linearised equations for the velocity field now yield
\begin{equation}
\frac{D \, \delta u}{Dt} = - \frac{\delta M}{r^2} - \bar{H} \, \delta u
\end{equation}
and
\begin{equation} 
\frac{D}{Dt} \left(\frac{D \, \delta u}{Dt}\right) + 3 \bar{H} \frac{D
\,\delta u}{Dt} - \bar{H}^2 \, \delta u = 0,
\label{lin2}
\end{equation}
where we have assumed that the background is described by a spatially
flat, $\Lambda=0$ cosmology.  Equation~\r{lin2} shows that the
evolution of the density perturbation down the streamline is
determined entirely by the background cosmology, and not by the local
properties of the perturbation.  It follows that there is nothing
special about the assumption of spherical symmetry and
equation~\r{lin2} is a general result for the growth of a velocity
perturbation in a dust-filled cosmology.

We can parameterise $\delta u$ in terms of a streamline starting point
and cosmic time $t$.  In this way the streamline derivatives are
replaced by derivatives with respect to $t$, and we arrive at the
equation
\begin{equation}
\delta \ddot{u} + 3\bar{H} \, \delta \dot{u} - \bar{H}^2 \, \delta u =
0. 
\end{equation}
Since $\bar{H}=2/(3t)$ we see that $\delta u$ has modes going as
$t^{1/3}$ and $t^{-4/3}$.  Only the former of these is growing with
respect to the background value.  If we assume that we are in an epoch
when the decaying mode can be ignored, we find that
\begin{equation}
\frac{D \delta u}{Dt} = \frac{1}{2} \bar{H} \delta u.
\end{equation}
It follows that the density and velocity perturbations in the linear
regime, when no decaying modes are present, are related by
\begin{equation}
\frac{3}{2} \bar{H} \delta u = -\frac{\delta M}{r^2}.
\label{lindenv}
\end{equation}
This provides a constraint on the initial conditions for our
models.

\section{Swiss Cheese Models}
\label{S-SCM}

The most popular method to date for theoretical calculations of CMB
anisotropies in the nonlinear regime has been to employ `Swiss Cheese'
models~\cite{rees68,nott82}.  In these models an overdense collapsing
region is surrounded by a compensating vacuum region which then
matches onto the external Universe.  Nottale~\shortcite{nott82}
developed a complicated set of equations for describing photon
redshifts in such a model.  The interior region is modelled by a
collapsing $\Omega >1$ cosmology, and three coordinate systems are
employed with two sets of matching equations at the boundaries.
Initial data consists of the set $\{H_0, \rho_0, H_c, \rho_c, r_i, z_c
\}$, where $H_0$ and $\rho_0$ are the values of the Hubble constant
and density at the present epoch, $H_c$ and $\rho_c$ are the
equivalent variables for the central cluster region as the photon
passes through the centre, $z_c$ is the redshift of a photon emitted
from the centre of the cluster at the epoch described by $H_c$ and
$\rho_c$, and $r_i$ is the cluster radius when the photon exits the
central core.  One of the five dimensionful quantities is arbitrary
and can be scaled to unity, so Nottale's setup defines a
five-parameter model.  To analyse this it is convenient to introduce a
set of dimensionless variables.  Along with $z_c$ these consist of the
two deceleration parameters $q_c$ and $q_0$, together with
\begin{equation} 
y \equiv H_c r_i \quad \mbox{and} \quad h \equiv \frac{H_c}{H_0}.
\end{equation} 
It is also convenient to define
\begin{equation}
n = \left(\frac{\rho_c}{\rho_0} \right)^{1/3} =  \left(\frac{q_c}{q_0}
\right)^{1/3} h^{2/3}.
\end{equation}

From the initial data it is possible to calculate the redshift of a
photon as it reemerges into the exterior Universe given its redshift
at the centre $z_c$.  The method relies on the fact that the initial
data define the epoch when the photon is at the cluster centre.  First
$H_c$, $\rho_c$, $r_i$ and $z_c$ are used to calculate the redshift of
the photon at the edge of the collapsing region.  This step is a
straightforward application of the Mattig relation~\cite[Chapter
13]{peeb-cosm}.  Next one computes the photon redshift after it
crosses the void region and enters the exterior Universe.  This
calculation is possible because of the existence of a timelike Killing
vector in the vacuum region, which ensures that $\omega(\Gamma + u)$
is constant as the photon crosses the vacuum.  From this it is
possible to calculate the redshift of the photon as it reenters the
unperturbed Universe as a function of the input variables.  A similar,
more involved calculation is performed in the opposite time direction
to obtain the total photon redshift across the perturbed region.  This
redshift is then compared with that of a photon coming from an
unperturbed region of the Universe to obtain a central temperature
decrement.

The resultant equations are complicated and cannot be solved
analytically.  Nottale~\shortcite{nott82} obtained a power series
solution by expanding in both $y$ (which is necessarily small) and
$z_c$.  A better approach, however, is to obtain an expression valid
for general $z_c$ using only an expansion in $y$.  This calculation is
tricky, but with the help of the symbolic algebra package Maple it is
possible to perform such an expansion.  The result, which we give here
for the first time, is
\begin{eqnarray}
\frac{\Delta T}{T}
&=& -4 q_0 \left(\frac{ny}{h}\right)^3 \Bigl\{ (1+ z_c)
(1+ 2 q_o z_c)^{1/2} \nn \\
& & \times
\bigl[ 1-\ln \left(\frac{n}{1+z_c} \right) \bigr]
-h \Bigr\} .
\label{Nottres}
\end{eqnarray}

\begin{figure}
\begin{center}
\epsfig{figure=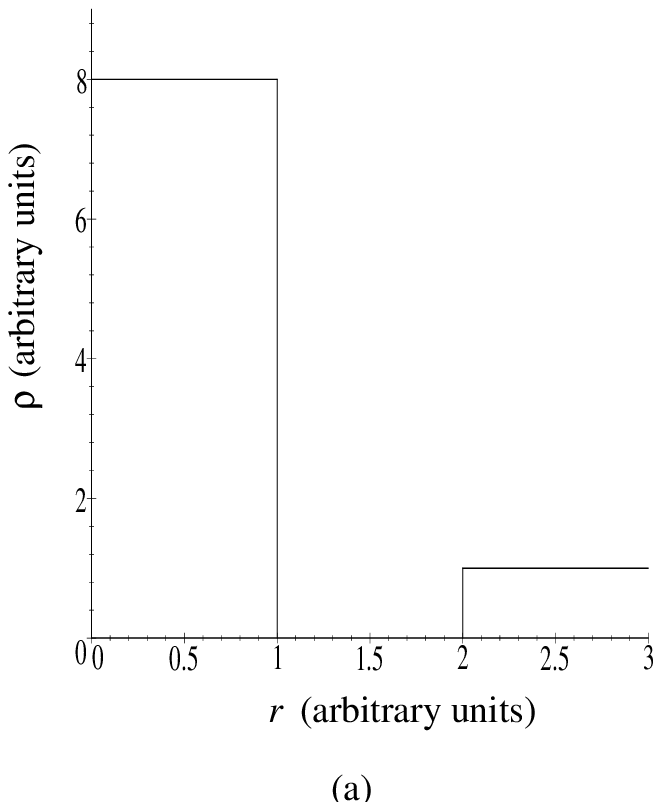,width=6cm}
\epsfig{figure=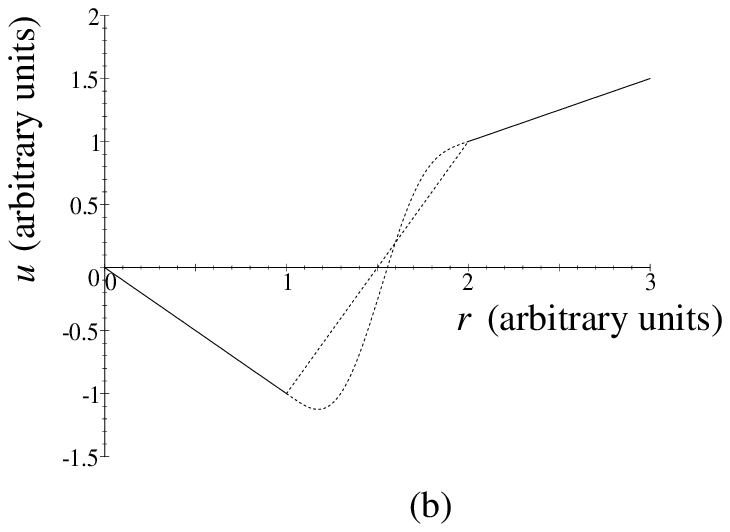,width=6cm}
\end{center}
\caption{{\sl Initial Data for Swiss Cheese Models}. (a) shows the
initial density distribution, with a uniform central overdensity and a
compensating vacuum.  The width of the compensating region ensures
that the total mass enclosed is the same as would have been enclosed
if the density was uniformly distributed with the exterior value.
(b) shows the initial velocity distribution.  The central region is
modelled by a collapsing $\Omega>1$ Universe, and the exterior region by
an arbitrary, expanding model.  There is no unique curve extrapolating
the initial velocity distribution across the vacuum.  Two different
curves are shown, which would produce two different models.}
\label{F.Ch1}
\end{figure}

That one can derive an equation such as~\r{Nottres} is somewhat
surprising.  From the Newtonian gauge perspective, the initial data
appears to under-determine the system.  This is illustrated in
Figure~\ref{F.Ch1}, which shows that the velocity profile is not
uniquely specified since various curves can be fitted in the vacuum
region.  Different curves produce different crossing times for photons
to cross the vacuum region and so produce different physical models.
The trick which makes the derivation of~\r{Nottres} possible is that
$z_c$ is specified as part of the initial data.  This effectively
places an extra constraint on the form of the velocity function in the
vacuum and, while not pinning down the form of the curve uniquely,
this does render the predictions unique.  The residual freedom in the
choice of velocity function amounts to a gauge freedom so has no
physical significance.

While the result~\r{Nottres} provides a useful estimate of the effect
of a cluster, this type of Swiss Cheese model has a number of
weaknesses.  The density distribution is highly unrealistic and must
limit the accuracy of any predictions from such models.  The same can
be said of the the high infall velocities in such models, which appear
to overestimate the CMB temperature decrement.  A further weakness is
the number of arbitrary parameters in the model.  While the cluster
width, density and redshift are certainly sensible parameters for such
a model, the velocity in the form of $H_c$ or $h$ is not an obvious
physical parameter.  Nottale addressed this weakness in a later
paper~\cite{nott84} by introducing the additional restriction that the
two matter regions issued from the same singularity.  This correlates
the (Newtonian) times in the internal and external regions, and allows
one to determine $h$ in terms of the remaining dimensionless
parameters.  This assumption is partially supported by Silk's result
that the effect of eliminating the decaying modes is to set the time
that each streamline leaves the initial singularity to a constant
value~\cite{silk77}.  This result only holds in the linear theory,
however, so it is not obviously applicable to Swiss Cheese models.  In
the following section we propose an alternative model which addresses
these weaknesses in a simple and elegant manner.

\section{Simple, Four-Parameter Models}
\label{S-MOD}

The control over the initial conditions afforded by the Newtonian
gauge approach provides for many models of the development of a
perturbation away from a homogeneous cosmology.  Here we discuss a
family of simple four-parameter models based on polynomial
perturbations in the density and velocity fields.  The perturbation is
of finite extent and the density distribution is compensated.  The
region outside the perturbation therefore evolves as a homogeneous
cosmology, and placing observers in this region allows for unambiguous
calculations of the CMB perturbation caused by the cluster.  The
perturbation is assumed to have grown from primordial fluctuations in
the very early universe, and to still be well described by the
linearised Einstein equations.  We therefore expect that the decaying
modes are negligible at the epoch when we specify the initial
conditions.  This physical requirement is imposed by ensuring that 
the initial velocity and density profiles satisfy
equation~\r{lindenv}.

\begin{figure}
\begin{center}
\epsfig{figure=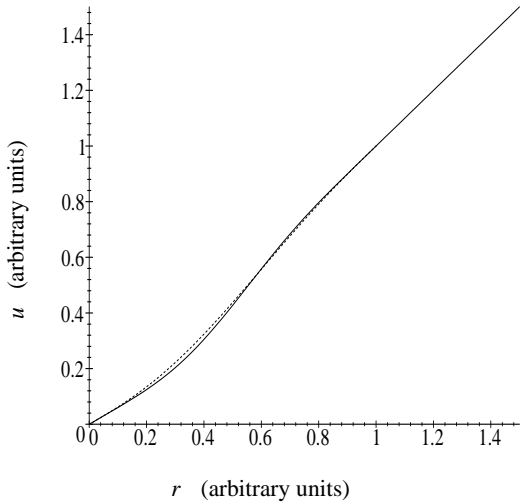,width=8cm}
\epsfig{figure=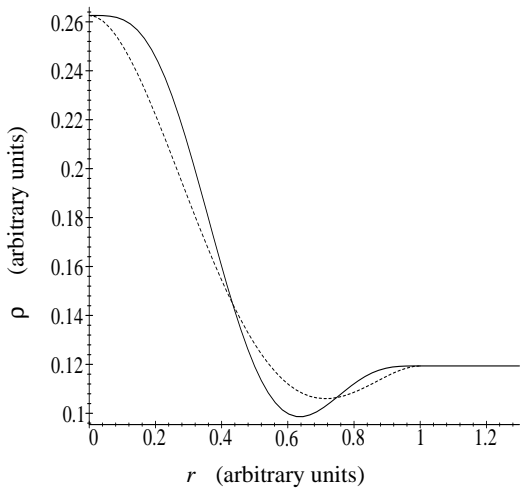,width=8cm}
\end{center}
\caption{{\sl Initial data for the 4-parameter model}.  The solid curves are
for $m=4$ and the dashed curves for $m=2$.  The
perturbation has radius $r_i=1$ (arbitrary units), beyond which the
velocity $u(t_i,r)$ is proportional to $r$ and the density is
uniform.  The perturbation is defined by the velocity gradient at the
origin (set to 0.6 here) and the radius $r_i$.  The density profile is
compensated, so the total mass enclosed in $r<r_i$ is the same as
would have been enclosed if the initial density was uniform.}  
\label{Fig2}
\end{figure}

In our model the velocity perturbation is controlled by two physical
parameters --- one specifying the width of the perturbation, and the
other the velocity gradient at the origin.  For models of cluster
formation the velocity gradient at the origin is slightly less than
that of the unperturbed Universe.  Voids can be modelled with a
slightly greater velocity gradient.  A third parameter $m$ controls
the degree of the polynomial describing the perturbation, and is chosen
to to ensure a realistic density profile in the nonlinear region.  The
polynomial describing the velocity perturbation has degree $2m+1$ and
is formed as follows.  At the centre we have $u=0$ and the velocity
gradient is given.  All other derivatives up to order $m$ are set to
zero.  At the edge we match $u$ and its first $m$ derivatives to the
exterior value, $u=rH_i$.  These boundary conditions specify a unique
curve for each value of $m$.  Curves with $m$=2 and 4 are shown in
Figure~\ref{Fig2}.

With the velocity profile determined, the density profile is found
from equation~\r{lindenv}.  The initial data for the exterior universe
consists of a value for the Hubble function $H_i$ and the density
$\rho_i$.  With the perturbation width given by $r_i$, we can introduce
the dimensionless variables
\begin{equation}  
x = \frac{r}{r_i}, \quad v(x) = \frac{u(r,t_i)-rH_i}{r_i H_i}, \quad
f(x) = \frac{\rho(r,t_i) - \rho_i}{\rho_i}. 
\end{equation}
In terms of these, equation~\r{lindenv} takes the form
\begin{equation}
x^2 f(x) = - \frac{d}{dx} (x^2 v(x)).
\end{equation}
Since $v(x)$ is a degree $2m+1$ polynomial, $f(x)$ is a polynomial of
degree $2m$.  Computing the density distribution in this manner
ensures that it is compensated at the boundary.  The external values
of $\rho_i$ and $H_i$, together with the magnitude and width of
the perturbation, form a set of four parameters, one of which is
arbitrary because of the freedom to rescale.  The initial density and
velocity perturbations are given by polynomials, so the function
$f(r_i)=\Gamma(r_i)^2-1$ also has a simple polynomial expression.  It
follows that all of the terms on the right-hand side of
equation~\r{denevol} can be calculated algebraically, enabling
accurate computation of the density on a given streamline.

\begin{figure}
\begin{center}
\epsfig{figure=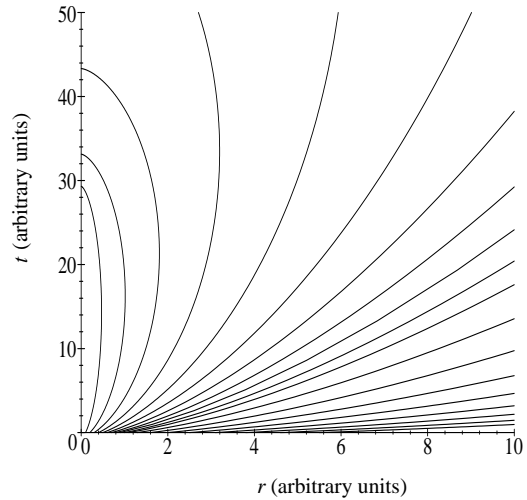,width=8cm}
\end{center}
\caption{{\em Matter streamlines for an $m=3$ model}.  The perturbation
has initial width 1, with $H_i=1$ and $\rho_i=3/(8\pi)$.  The velocity
gradient at the centre of the perturbation is 0.95.  The central
region is moving inwards relative to the Hubble flow, so recollapses
to a singularity in a finite time.  The form of the initial data
ensures that streamline crossing does not occur.}
\label{Fig3}
\end{figure}

\begin{figure*}
\begin{center}
\epsfig{figure=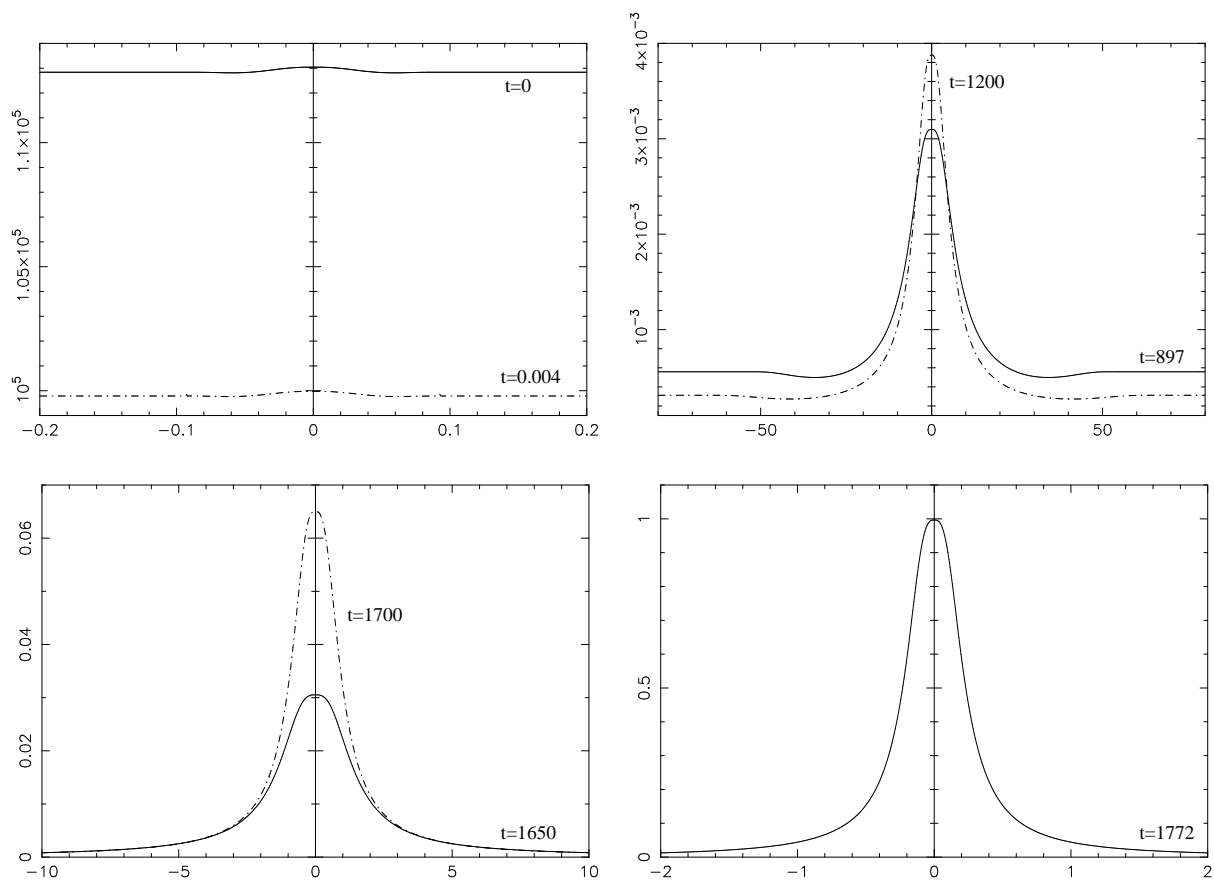,width=16cm}
\end{center}
\caption[dummy]{{\em Density evolution for an $m=3$ model}.  The
initial density profile contains a small, compensated perturbation
around the background level.  Initially, the density falls with the
expansion of the Universe.  After $t=897$ the streamlines turn round,
the central region starts to collapse and the density there increases.
The perturbation evolves to give the final density distribution shown
at $t=1772$.  This closely matches the observed density profiles in
galaxy clusters~\cite{DDGHL-II}.  The units are arbitrary.}
\label{Fig5}
\end{figure*}

An example of the streamlines resulting from such an initial setup is
shown in Figure~\ref{Fig3}.  The perturbed region has a slight
undervelocity, so after a finite time it starts to recollapse and form
a compact cluster.  The figure illustrates that streamline crossing
does not occur in our models.  An example of the density distribution
resulting from growing this perturbation is shown in
Figure~\ref{Fig5}.  This density distribution closely matches the King
profile observed in galaxy clusters~\cite{DDGHL-II}, a remarkable
result given the simplicity of the model.

The fact that the density distribution evolves to a realistic profile
makes it possible to choose input parameters to achieve realistic
simulations of the effect of a cluster.  A photon can then be sent
through the cluster to simulate the effect on the CMB.  The epoch at
which the photon enters the cluster region is the fourth physical
parameter in the model.  One can then perform detailed calculations of
the lensing properties of the cluster and its effect on the CMB.
These are contained in an accompanying paper~\cite{DDGHL-II}.

The main limitation with the model presented here is the lack of
pressure support.  This has two effects.  First, the infall velocities
in the cluster are unrealistically large compared to a virialised
cluster.  This will tend to overestimate the temperature decrement due
to the cluster, since it is the evolution of the velocity difference
$\Delta(t,r)$ with time which is central in determining the magnitude
of the perturbation.  The second effect is that a central singularity
forms too soon, when the first streamline reaches the origin.  There
are no difficulties in extending the model beyond this time, provided
the photon has exited the cluster before the horizon has formed, but
the initial conditions must ensure that singularities do not form at
early epochs.  It turns out that introducing the perturbation at
recombination does not pose any problems numerically, and we typically
set our initial conditions at $z=1000$.

\section{Conclusions}

We have shown that both spherically symmetric collapse and cosmology
can be studied in a single global coordinate system, which we have
called the Newtonian gauge.  This gauge considerably simplifies the
task of modelling cluster formation and its effects on the CMB.
Observers can be placed in a region of the model corresponding to an
FRW Universe, and both the lensing and CMB temperature decrement are
easily modelled.  In an accompanying paper~\cite{DDGHL-II} detailed
computations are performed with this model and compared with other
work and with observations.  Numerical integration along photon paths
are performed to compute the lensing effects of the cluster.  The
effects on the CMB power spectrum are also computed by evolving a
simulated background (with Gaussian fluctuations) through the cluster
to the observer.

A study of Swiss Cheese models within the Newtonian gauge revealed the
surprising result that the models only yield unique predictions
because of the particular way the initial data is given.  A new
formula for the CMB temperature decrement in a Swiss Cheese model was
presented, and it was argued that the number of arbitrary parameters
in such models limits their usefulness.

The problems with Swiss Cheese models are overcome in a simplified
model in which a small density and velocity perturbation is introduced
at recombination and allowed to grow to give a compensated density
profile.  The centre of this density distribution models a cluster
with a density profile close to that of the King profile.  Photons can
then be sent through this density distribution to model the effects of
clusters at arbitrary redshifts.  The lack of pressure support means
that the model probably over-estimates the perturbations induced in
the CMB.  This will certainly be the case for nearby clusters which
are virialised and do not have large infall velocities.  Recent work
on structure formation, however, has suggested that clusters with
large infall velocities could well have formed around redshifts of 2
and higher.  If so, the model presented in this paper could prove very
useful for calculating the effects of such clusters.

\section*{Acknowledgements}

We thank an anonymous referee for useful comments.  CD thanks the
Lloyd's of London Tercentenary Foundation for their financial support.
MPH thanks Trinity Hall, Cambridge, for their support in the form of a
research fellowship.


\appendix

\section{Local Observables}
\label{appA}

It is instructive to see how one might measure $\Gamma$, $r$ and $u$ in
the collapsing region with only local measurements of the
gravitational field, and to see how the method fails outside the
cluster.  The key variable is the strength of the radial tidal force.
For two objects a (small) proper distance $d$ apart the relative
acceleration of one from the other is
\begin{equation}
a = d \left( 4\pi \rho - \frac{2M}{r^3} \right).
\end{equation}
By measuring the force necessary to prevent this tidal acceleration and
comparing this with the local density it is possible to calculate the
local value of $M/r^3$.  Of course, in a homogeneous region this would
be just  $4\pi\rho/3$.  

Next we measure the variation of $M/r^3$ with time for freely-falling
observers.  This goes as
\begin{equation}
\frac{D}{Dt} \left(\frac{M}{r^3} \right) = -\frac{3M}{r^3} \frac{u}{r},
\end{equation}
so measurements of this lead to a determination of $u/r$.  It follows
from the definition of $\Gamma$~\r{defg1} that the value of
$(\Gamma^2-1)/r^2$ can also now be determined.  Finally, we need to
measure the local variation of $M/r^3$ with proper distance on a
constant timeslice.  This goes as
\begin{equation}
\Gamma \frac{\partial}{\partial r}  \left(\frac{M}{r^3} \right) = \left(
\frac{4}{3} \pi \rho - \frac{M}{r^3} \right) 3 \frac{\Gamma}{r}.
\label{obs2}
\end{equation}
From this we can extract $\Gamma/r$, and from all of the above measurements
we can reconstruct $\Gamma$, $r$ and $u$.

In the homogeneous region, however, this final step is not possible
because the right-hand side of~\r{obs2} vanishes.  This prevents one
from determining $r$ from local measurements alone, which is
reassuring because in the homogeneous region all local observables
should be functions of time only.

\label{lastpage}

\end{document}